\documentclass[manuscript]{aastex631}
\usepackage[utf8]{inputenc}
\usepackage{times}
\usepackage{graphicx}
\usepackage{rotating}
\usepackage{wrapfig}
\usepackage{xcolor}

\newcommand{\jyb}{Jy~beam$^{-1}$}
\newcommand{\kms}{km~s$^{-1}$}
\newcommand{\nh}{N$_2$H$^+$}
\newcommand{\meth}{CH$_3$OH}

\newcommand{\sgra}{Sgr~A$^*$}

\newcommand{\aproxi}{$\sim$}

\begin{document}
\title{Parsec-scale SiO emission in the Galactic Center}

\author[0000-0002-2035-0644]{N. Wenner}
\affiliation{Department of Physics and Astronomy,
Northwestern University,
2145 Sheridan Rd,
Evanston, IL, 60208,
USA}
\affiliation{Center for Interdisciplinary Exploration and Research in Astrophysics (CIERA),
Northwestern University,
1800 Sherman Ave,
Evanston, IL, 60201,
USA}

\author[0000-0002-7570-5596]{C. J. Chandler}
\affiliation{National Radio Astronomy Observatory, P.O. Box O, Socorro, NM, 87801, USA}

\author[0000-0003-3503-3446]{J. M. Michail}
\affiliation{Center for Astrophysics $|$ Harvard \& Smithsonian, 60 Garden Street, Cambridge, MA 02138-1516; USA}
\affiliation{NSF Astronomy and Astrophysics Postdoctoral Fellow}

\author[0000-0001-9300-354X]{M. Gorski}
\affiliation{Center for Interdisciplinary Exploration and Research in Astrophysics (CIERA),
Northwestern University,
1800 Sherman Ave,
Evanston, IL, 60201,
USA}

\author{J. Braatz}
\affiliation{National Radio Astronomy Observatory, Charlottesville, VA 22903, USA}

\begin{abstract}
The central 5~parsecs of our Galactic Center is rich in various molecular tracers. The region is observed to be hot, highly ionized, and volatile, resulting in complex chemistry and kinematics. Countless molecular observations of the central region orbiting the central supermassive black hole \sgra\ have revealed a stable grouping of clouds with a disk-like structure known as the circumnuclear disk (CND) orbiting \sgra\ between 1.5 and 3~pc. The CND houses the densest population of gas outside the ionized streams in the minispiral falling onto \sgra. However, the small-scale kinematics of the clouds and gas are poorly understood in this region. In this paper, we report single dish SiO (1--0) observations of the central 5~pc surrounding \sgra\ taken with the Green Bank Telescope. Most of the SiO (1--0) traces the southern streamer and Sgr A East--neither of which are associated with the CND. However, one region in the CND, known as the NE Arm, shows high-intensity SiO emission unlike in the rest of the CND, indicating the NE Arm as a candidate region of either cloud-cloud collision or early star formation. Further high resolution observations of the region are needed to distinguish between the two scenarios.

\end{abstract}

\section{Introduction}

The kinematics within the central 5 parsecs of the Galactic Center (GC) are complicated due to the existence of the central supermassive black hole (SMBH) \sgra. Over 30 magnitudes of visual extinction lie between us and the central SMBH, making radio and submm observations critical in determining the kinematics of the gas orbiting the SMBH. Different wavelength regimes reveal different components of the Sgr A complex's structure,structure, such as the mini-spiral, which can be seen in radio. However, radio spectral information is required to reveal the dynamics of orbiting clouds observed in other wavelength regimes such as the infrared.

\begin{figure}[ht]
    \centering
    \includegraphics[width=12cm]{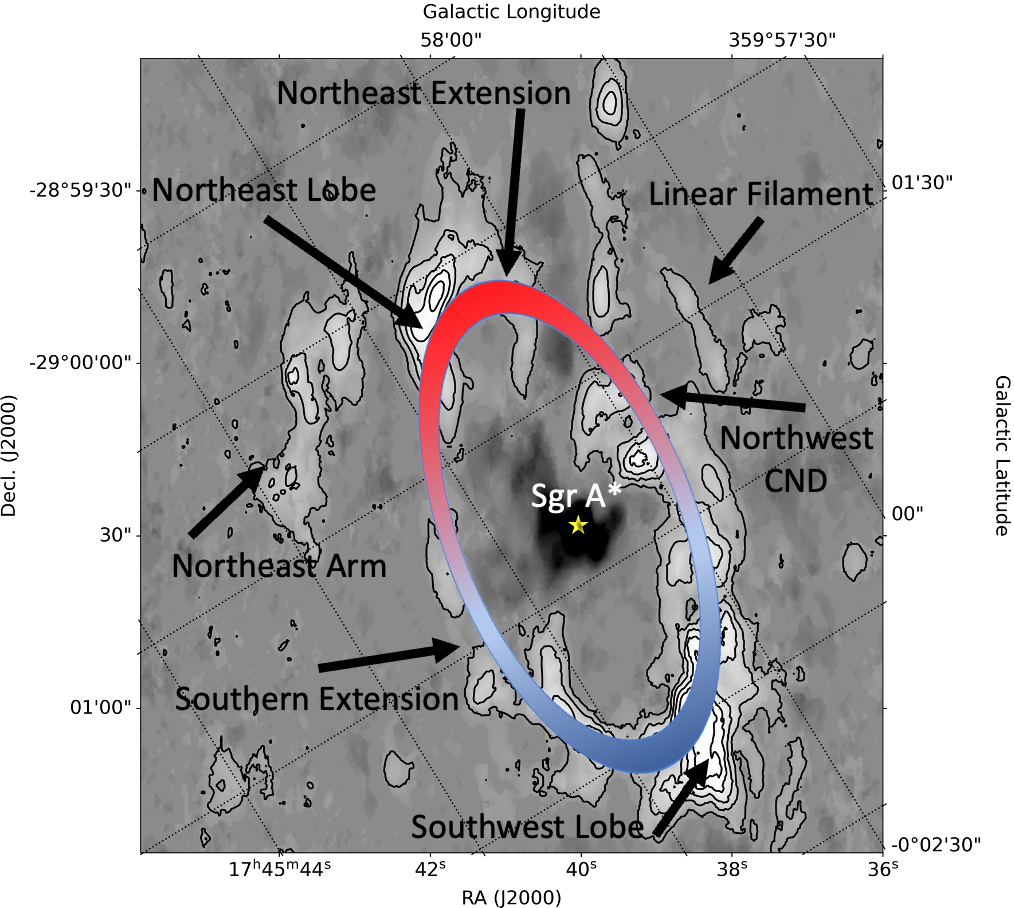}
    \caption{Map of the CND displaying the inclination and structure of the ring as derived by \citet{Tsuboi2018ALMACenter}, which agrees with the model based on infrared observations of \citet{Lau_2013_CNR}. The ring is drawn to scale to match the model of \citet{Lau_2013_CNR}. The contours are of integrated HCN (1—0) emission from \citet{Christopher2005HCNDISK}. The blueshifted regions lie to the southwest, and the redshifted regions are to the northeast. \sgra's location is marked by a yellow star. Galactic coordinates also plotted, showing how the CND is oriented relative to the Galactic Plane. Features of the CND are labeled  following the naming conventions used by \citet{Montero-Castano2009GasDisk}.}
    \label{fig:HCN_map}
\end{figure}

One such structure that appears in the radio molecular lines but not radio continuum is known as the circumnuclear disk (CND), traced directly by spectral lines of HCN, CS, NH3, CN, and various other molecules that are typically good at tracing dense gas in addition to IR continuum \citep[e.g.][]{Coil2000TheCenter,Christopher2005HCNDISK,Martin2012SurvivingSagittarius,Lau_2013_CNR,Hsieh2021TheCenter}. The CND orbits between 1.5 and 3~pc around \sgra\ \citep{Christopher2005HCNDISK}. The formation and stability of the CND are topics of debate and active study \citep[e.g.][]{Christopher2005HCNDISK,  Montero-Castano2009GasDisk,Martin2012SurvivingSagittarius, Tsuboi2018ALMACenter, Sormani2020SimulationsFormation} as the gas densities in this region are also hotly debated \citep[see][for a review]{Bryant2021TheCentre}. The rotational velocity of the CND is estimated to be $115 \pm 10$~\kms\ with the disk plane inclined at an angle of $30\degr$ to the line of sight \citep{Tsuboi2018ALMACenter}. Figure \ref{fig:HCN_map} shows the orientation of the CND with HCN (1--0) \citep{Christopher2005HCNDISK} and a colored ring to display redshifted vs blushifted components. The CND is divided into sub-structures within the disk, and this paper will use the naming conventions from \citet[][reproduced in Figure \ref{fig:HCN_map}]{Montero-Castano2009GasDisk}.
Spectral lines provide a fuller picture of the mechanisms driving the evolution of gas and dust near the SMBH; however, the dense gas tracers noted above only reveal a subset of the kinematics of the region. Therefore, tracers of other kinematic features, such as shocks, are necessary to explore the region in more depth. Understanding the kinematics in the CND will aid in the pursuit to understand the effects SMBHs have on their environment, such as whether they quench or induce star formation.

SiO (1--0) differs from other molecular tracers such as HCN and CN since silicon primarily resides embedded within dust grains, thus hus it has a very low abundance in the interstellar medium \citep[$\lesssim$ $10^{-11}$ relative to H$_2$;][]{Ziurys1989INTERSTELLARCHEMISTRY}. In molecular clouds, SiO forms in the gas phase when the silicon is released through sputtering or sublimation of the dust grains in shocks \citep{Gusdorf2008SiOWaves}, and it subsequently freezes out again on timescales of order 3000 years \citep{Guillet2011ShocksShocks}. The presence of SiO therefore uniquely traces recently-shocked gas, which can occur as a result of star formation processes and cloud-cloud collisions \citep{Zeng2020Cloud-cloudClouds,Jrgensen2020AstrochemistryStars}. SiO emission traces outflows associated with protostars as an unambiguous sign of star formation \citep{Bally2016ProtostellarOutflows} in the form of high velocity shocks with bright, broad ($> 8$~\kms), and potentially winged spectral lines \citep[e.g.][]{Lopez-Sepulcre2016THE2264-C,Kim2023ARegions}. Bipolar outflows are early signs of star formation and appear before the formation of ultra-compact HII (UC HII) regions \citep[][and references therein]{Motte2018High-MassWay}, making SiO an excellent tracer of high-mass star formation in the CND. On the other hand, SiO also traces other shocked regions where star formation is not occurring such as regions of cloud-cloud collisions. Shocks in regions of cloud-cloud collisions are low velocity with narrower line widths than observed outflows  (\aproxi1–4~\kms) and also show bridging features that appear spatially where the two velocities of the colliding clouds are interacting  \citep[e.g.][]{Zeng2020Cloud-cloudClouds,Cosentino2020SiOClouds,Kim2023ARegions}. Regions of cloud-cloud collisions can potentially induce star formation but may not actively be producing new stars. Cloud-cloud collisions also occur on larger scales than outflows, so shocks from outflows are more beam diluted than cloud-cloud shocks when observed with single-dish telescopes. Therefore, determining the difference between the two spectra are important to determine their kinematics.
Understanding how stars form in the CND is a poorly understood but important topic, as a cluster of O/B stars exists orbiting around Sgr A* with an unknown origin \citep{Paumard2006THE1} but likely originated in the CND \citep{Mapelli2016ModellingCentre,Dinh2021EffectsDisk}.

We present archival single dish Robert C. Byrd Green Bank Telescope (GBT) SiO (1--0) data covering the entirety of the CND for the first time in this paper. The observations and data reduction procedures are described in Section \ref{sec:Obs}. We report the results in Section \ref{sec:results}. In Section \ref{sec:Discuss}, we discuss the results and compare them to higher transitions of SiO observed in the CND, including what the (1--0) transition can tell us about the star formation potential within the CND. We state our conclusions in Section \ref{sec:conclusion}.

\subsection{Galactic Center Dynamics}

\begin{figure}[ht]
    \centering
    \includegraphics[width=12cm]{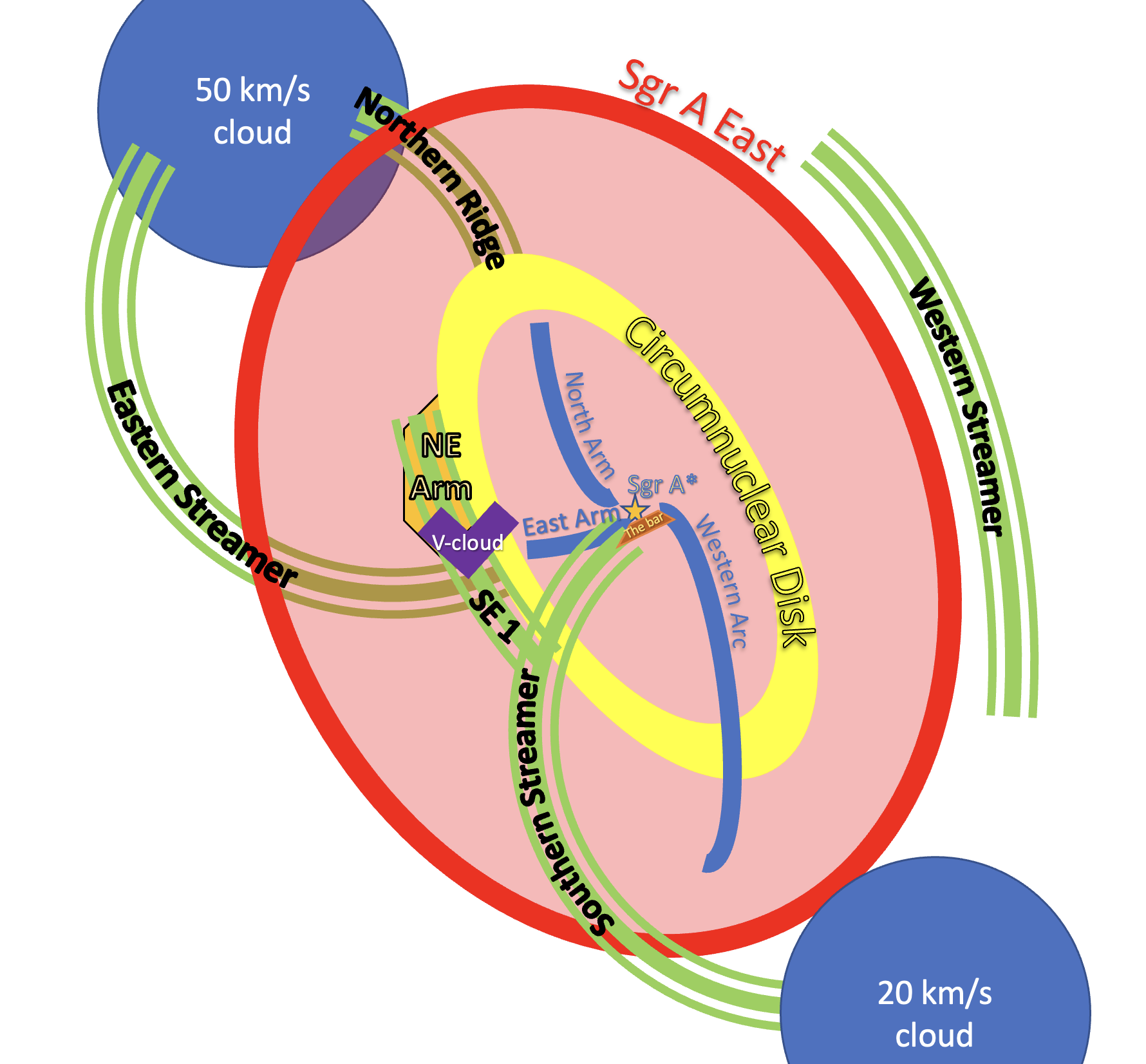}
    \caption{Cartoon layout of the main structures in the vicinity of \sgra\ inspired by Figure 36 from \citet{Bryant2021TheCentre}. Various locations and features are approximate and illustrate the projected view of the Galactic Center in equatorial coordinates.}
    \label{fig:Galactic_Center_schematic}
\end{figure}

Beyond the central SMBH at the center of the Milky Way, there are many complex molecular clouds, streamers, and other features that impact the central dynamics. Figure \ref{fig:Galactic_Center_schematic} is a cartoon inspired by Figure 36 from \citet{Bryant2021TheCentre} that illustrates the approximate locations and projections of these features surrounding \sgra. At the center is \sgra, where the three arms of what is known as the minispiral bring ionized gas onto the SMBH. Those features are called the East Arm, North Arm, and Western Arc. The small feature, known as ``The Bar," is the most blueshifted of the minispiral features and situated between the Western Arc and East Arm projected south of \sgra. The minispiral features show up in continuum observations and are believed to be the primary mechanisms feeding mass onto \sgra\ \citep[e.g.,][and references therein]{Bryant2021TheCentre}. The velocities in the minispiral are wide-ranging. The East Arm has velocities  100--300~\kms, the North Arm has velocities 20--120~\kms, the Western Arc has velocities between -200 to 100~\kms, and the Bar has velocities between -250 and -200~\kms\ \citep{Zhao2010THEGALAXY}. Surrounding the minispiral is the molecular circumnuclear disk that orbits the SMBH \citep[e.g.,][]{Christopher2005HCNDISK,Montero-Castano2009GasDisk} with velocities ranging between $\pm$150\kms. In the foreground on the eastern edge of the CND is an infrared dark cloud, called the v-cloud \citep{Bergin2007ColdFormation,Moser2017ApproachingA}, having velocities between 10--30~\kms. Behind these features is Sgr A East, an expanding supernova remnant shell traced by water and methanol (\meth) masers \citep{Yusef-Zadeh2015CompactA,Yusef-Zadeh2015SIGNATURESA, McEwen2016CENTER}. The 20~\kms\ and 50~\kms\ molecular clouds are separated south and north of \sgra\ by about 10~pc respectively. The 20~\kms\ cloud is closer to the Sun along the line of sight than the \sgra\ complex, while the 50~\kms\ cloud is located behind the \sgra\ complex \citep[e.g.][and references therein]{Bryant2021TheCentre}. Several features called streamers bring gas and dust from the 20~\kms\ and 50~\kms\ molecular clouds into the central 5 parsecs surrounding \sgra, however, their interactions with the CND are poorly understood \citep[e.g.][]{Coil2000TheCenter, Bryant2021TheCentre}. The southern streamer has a secondary arm called SE1 that goes up towards the NE Arm of the CND where the v-cloud overlaps and is traced by ammonia \citep{Coil2000TheCenter}.

\section{Observations and Data Reduction} \label{sec:Obs}
We report previously unpublished SiO$_{v=0}$ (1--0) 
data for the central 5~pc of the Galactic Center centered on \sgra, obtained using the Green Bank Telescope (GBT).
The GBT observations were taken on 2008 May 11 as part of project 7B-037, using frequency-switched point mapping with a frequency throw of 2~MHz. A total of 110 pointings were taken to cover the full CND. An under-sampled map of the CND is constructed of 80 of these pointings as seen in Figure \ref{fig:CNDMAP_under}. Each of these observations spent 106 seconds on target. An additional 30 pointings covered the NE Arm of the CND with near-Nyquist sampling (see Figure \ref{fig:NEArm_sample}). The pointings covering the NE Arm spent 53 seconds on source. The details of the observation are listed in Table \ref{tab:vla_obs}. The spectrometer had a total bandwidth of 400~MHz and 12.207~kHz channels for both maps.

\begin{figure}
\fig{undersampled_CND.png}{0.8\textwidth}{}
\caption{HCN (1--0) intensity map of the CND showing the location of the 15\arcsec\ FWHM GBT pointings of the under-sampled map overlaid. Each number in the circle corresponds to the scan number. The corresponding SiO (1--0) spectra for each pointing are shown in the corresponding location of each pointing on the map in Figure \ref{fig:SiO_CNDMAP}. The location of \sgra\ is labeled with a yellow star.}

\label{fig:CNDMAP_under}
\end{figure}

\begin{figure}
\fig{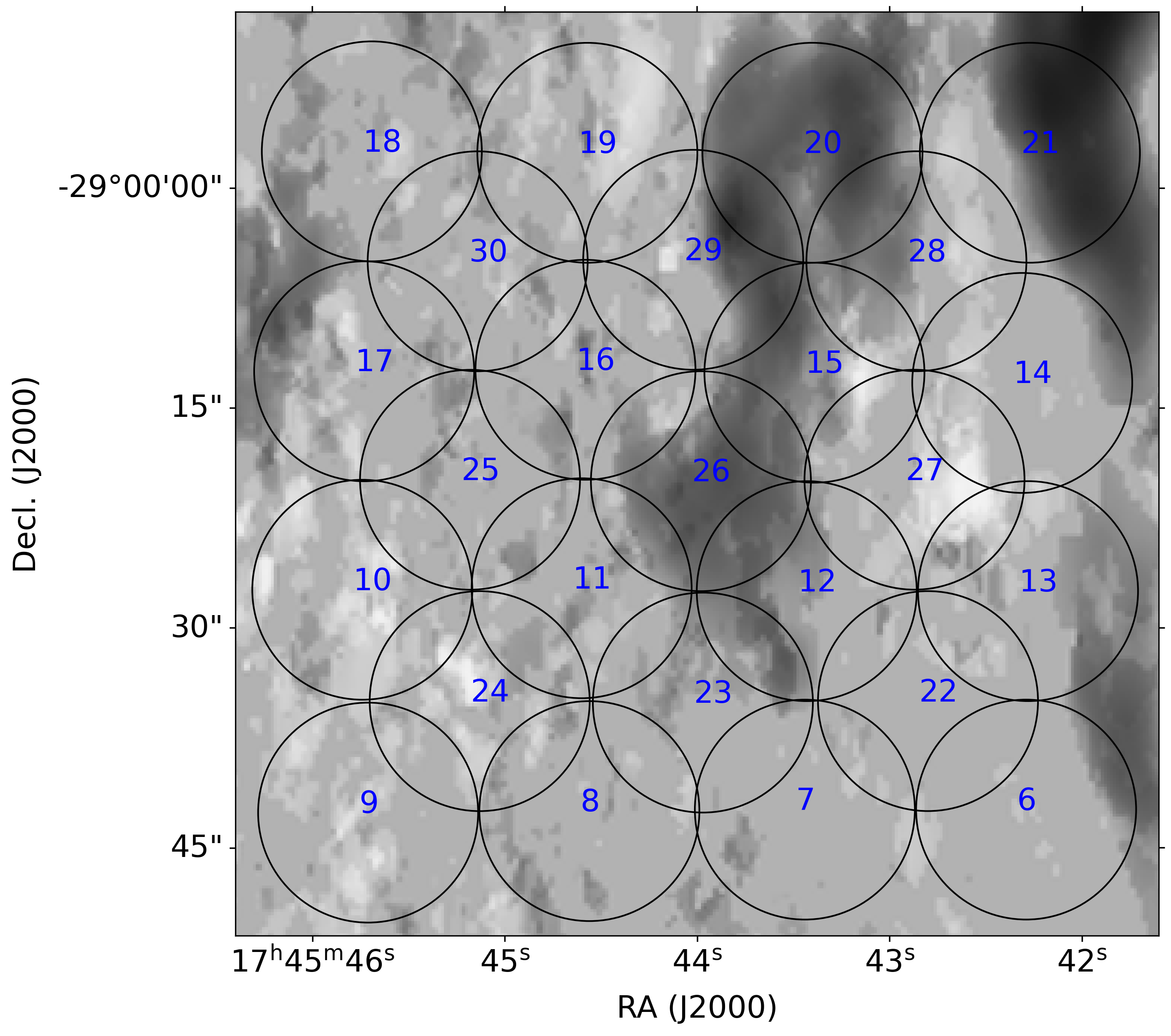}{0.8\textwidth}{}
\caption{HCN (1--0) intensity map of the CND showing the 15\arcsec\ FWHM GBT pointings of the more densely-sampled NE Arm map. Each number in the circle corresponds to the scan number in the original data. Each pointing spent 53 seconds on source.
}
\label{fig:NEArm_sample}
\end{figure}

\begin{figure}
\fig{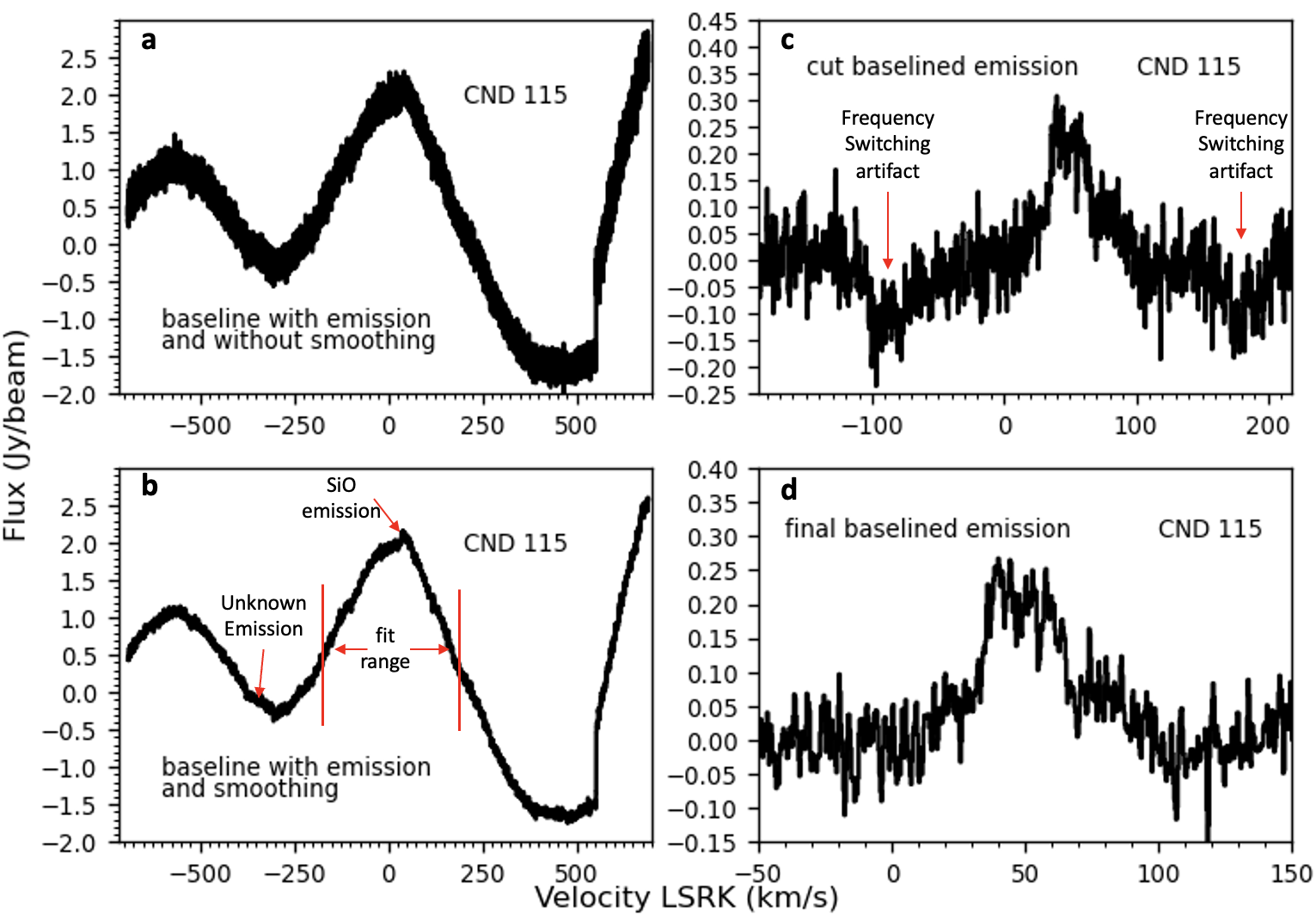}{.9\textwidth}{}
\caption{Spectral reduction example for SiO (1–0) data. Top Left: Raw GBT data prior to baseline subtraction. Bottom Left: same as top right after smoothing. The SiO emission is marked, as well as the unknown emission observed. The fit range for the baseline is also shown. Top Right: baseline-subtracted emission of SiO before cutting out the frequency-switching artifacts. Bottom Right: final spectrum of the SiO emission after all reductions are applied.
}
\label{fig:SiO_raw_data}
\end{figure}

Figure \ref{fig:SiO_raw_data} shows an example baseline fitting procedure for the SiO data. The observations were obtained when the GBT suffered from poor spectral baselines in Q-band \citep{Bradley2003AFront-End}, as can be seen in Figure \ref{fig:SiO_raw_data}a. Additionally, another weak unidentified spectral line appears around 43.426~GHz and is found in nearly every single spectra (see Figure \ref{fig:SiO_raw_data}b). These two problems, compounded with the narrow (\aproxi135~\kms) frequency-switching spacing, made constructing proper spectral baselines for the data complicated. Therefore, we focus on data between $v_{\rm LSR}= \pm 200$~\kms\ centered on the SiO (1--0) line at 43.423~GHz. This is the range of the kinematics of the gas within the CND, and the unidentified line is far beyond the expected range of the existing kinematic model of the CND, ruling out SiO (1–0). The typical polynomial fit to the data for baseline subtraction was 8th order. Lower-order polynomials failed to produce good fits. We applied 4-channel boxcar smoothing, decimated by a factor of 4, then one round of Hanning smoothing to the data before baseline fitting. After subtracting the baselines, artifacts from the frequency switching show clearly and are easily mistaken for absorption features if not removed (see Figure \ref{fig:SiO_raw_data}c for reference). These are easy to identify as they are $\sqrt{2}$ the strength of the emission at frequencies offset by the frequency-throw and have the same line shape as the emission \citep{Thomas2012GBTIDLGuide}. Therefore, we truncate the spectra to only include data between -75 and 190~\kms, which removed the frequency-switching artifacts while keeping the emission from the CND (see Figure \ref{fig:SiO_raw_data}d). In the end, the final reduced spectra resulted in 779 usable channels with bandwidth of \aproxi38~MHz and 48.82~kHz (0.337~\kms) channel widths.

We then used the cube-generating code \href{https://github.com/GreenBankObservatory/gbtgridder/tree/cygrid_dev}{gbtgridder}\footnote{gbtgridder is a beta-release of software for creating cubes from single dish data \url{https://github.com/GreenBankObservatory/gbtgridder/tree/cygrid_dev}} to produce a cube using all 110 pointings (see Figures \ref{fig:NEArm_sample} and \ref{fig:CNDMAP_under} for reference). We interpolated all the pointings onto a uniform 2\arcsec\ pixel grid using a 19.4\arcsec\ full width half max (FWHM) Gaussian, resulting in a 71x69 pixel spectral cube with an effective resolution of 25\arcsec. The final rms noise for the cube is 35~m\jyb. Additionally, we estimate the uncertainty in the flux scale to be about 20\% as is standard for the GBT at the time the data were taken \citep{Thomas2012GBTIDLGuide}. Finally, we produced moment maps with a mask for values below 2~$\sigma$ (\aproxi0.07~\jyb) to avoid noise contamination.

\begin{table}
    \centering
    \caption{Parameters for GBT 2008 Observations}
    \begin{tabular}{l l}
    \hline \hline
          Parameter & All Pointings \\ \hline
         
          Date & 2008 May 11 \\
          GBT Band & Q \\
          Molecule & SiO (1--0)$_{v = 0}$ \\
          Rest Frequency & 43.423 GHz \\
          Total bandwidth (MHz) & 400 \\
          Total number of pointings & 110 \\
          Total number of pointings full CND & 80 \\
          Total number of pointings on NE Arm & 30 \\
          No. of channels per spectral window  &  16384 \\
          Channel width & 0.337 \kms\ (0.048~MHz) \\
          RMS noise per pointing & 40~m\jyb\ channel$^{-1}$ \\
          FWHM of beam (arcsecond) &   15 \\
          Jy/K conversion factor & 0.24 \\

    \hline \hline
        Parameter & Cube \\ \hline

        No. of channels & 779 \\
        Channel width & 0.337~\kms \\
        RMS noise & 29~m\jyb\ channel$^{-1}$ \\
        FWHM of beam (arcsecond) & 25 \\
        Jy/K conversion factor & 0.67 \\

\hline \hline
  
    \end{tabular}
    \label{tab:vla_obs}
\end{table}

\section{Results} \label{sec:results}

Section \ref{sec:results} is divided into several parts. The first describes the individual pointings in the under-sampled CND map. The second presents the interpolated cubed data with both the Nyquist-sampled NE Arm and the under-sampled CND map.

\subsection{SiO (1--0) Postage Stamp Map} \label{subsec:resultsPointings}

\begin{figure}

\fig{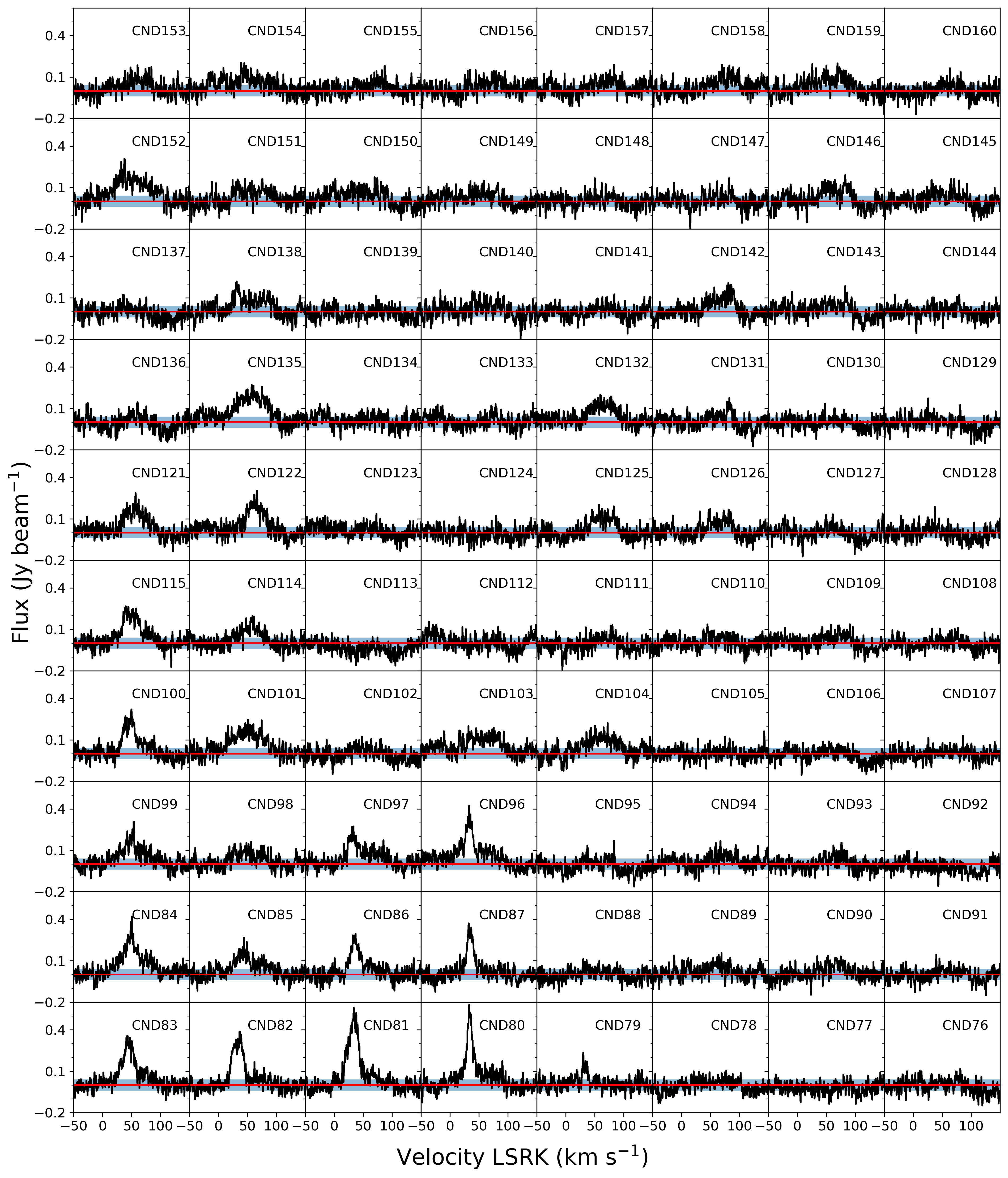}{0.9\textwidth}{}
\caption{Spectra showing SiO emission in the CND map are shown in the corresponding location of each pointing on the map in Fig. \ref{fig:CNDMAP_under}. The y-axis shows the flux in units of \jyb. The x-axis shows the central velocity of the SiO emission. The red line is the zero line. The blue shading represents the rms noise at 1$\sigma$ (0.04~\jyb). The majority of the SiO emission is concentrated south-east of the CND and also in the NE Arm.
}
\label{fig:SiO_CNDMAP}

\end{figure}

\begin{figure}

\fig{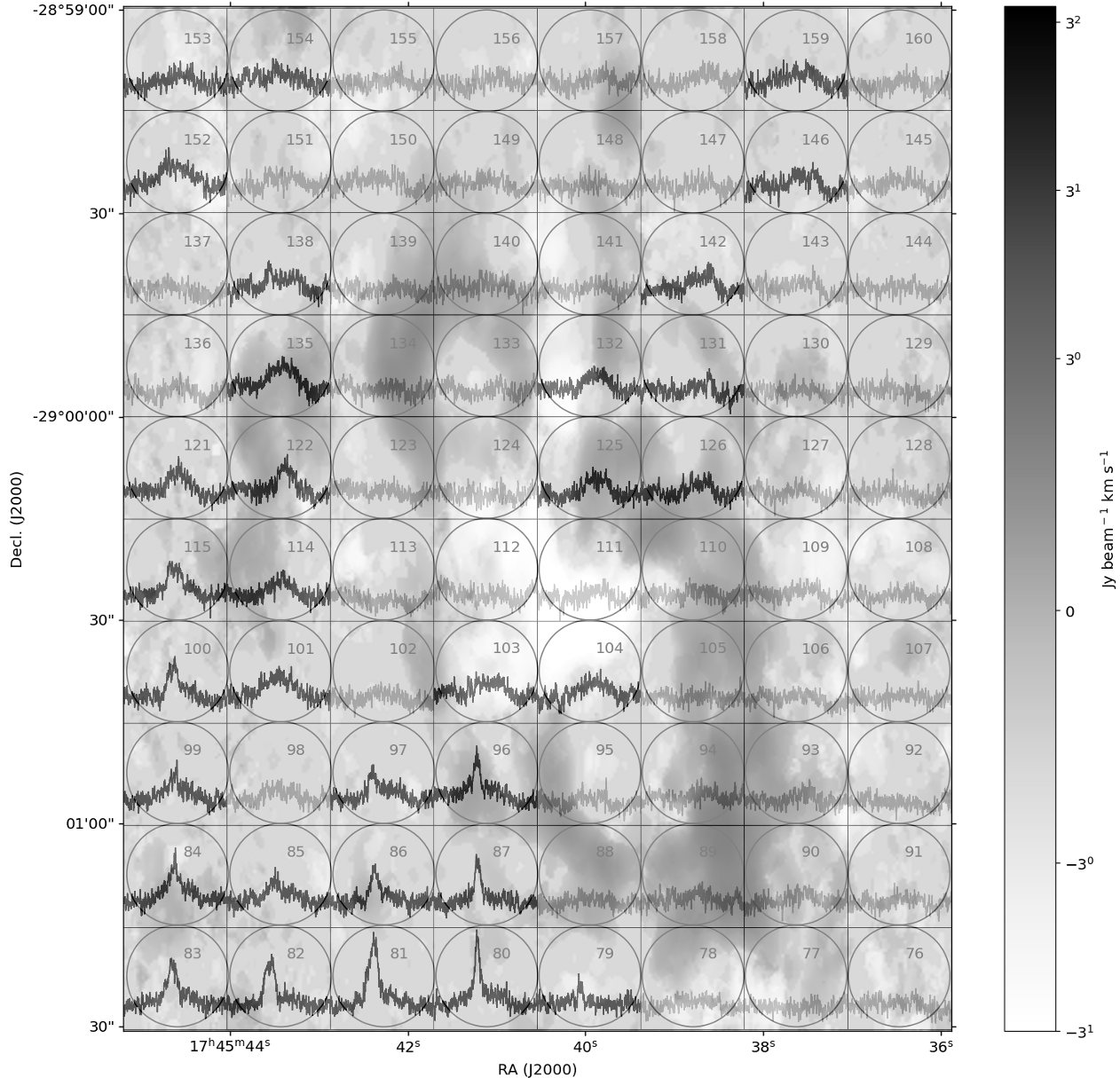}{0.9\textwidth}{}
\caption{Spectra showing SiO emission in the CND overlayed on an intensity map of HCN are shown in the corresponding location of each pointing encircled in the 15\arcsec\ beam. The y-axis shows the flux in units of \jyb. The x-axis shows the central velocity of the SiO emission. The emission over 3~$\sigma$ are highlighted with dark lines, while those that are not are plotted in gray. The majority of the SiO emission is concentrated south-east of the CND and also in the NE Arm
}

\label{fig:SiO_CNDMAP_HCN}

\end{figure}

A total of 80 spectra covering the central 5~pc around \sgra\ are reported here. Out of the 80 spectra, only 34 have emission greater than 2$\sigma$.
Figure \ref{fig:SiO_CNDMAP} shows a postage stamp map of all the spectra located at their associated pointing locations from Figure \ref{fig:CNDMAP_under}. The x-axis ranges from -50 to 150~\kms, and the y-axis ranges from -0.2 to 0.6~\jyb. A red line marks the zero-velocity line for clarity. All SiO emission detected falls within the ranges stated above.
The central velocities range from 20--90~\kms\, and linewidths (FWHM) range between 10--80~\kms. These ranges are consistent with structure both in and around the CND. Figure \ref{fig:SiO_CNDMAP_HCN} shows the same spectra from Figure \ref{fig:SiO_CNDMAP} plotted over an HCN map to better illustrate the physical location of the emission. The faded spectra are those with no emission above a 3~$\sigma$ threshold.

The strongest emission appears southeast of the CND. The peaks are sharp and narrower than that of ambient processes, indicative of shocks \citep{Gusdorf2008SiOWaves,Kim2023ARegions}. The peak velocities of the structures to the SE range between 30 and 45~\kms. The velocities and locations of the peaks here are consistent with the southern streamer traced by ammonia \citep{Coil2000TheCenter, Mcgary2001INCLOUDS} that is believed to bring material up from the 20~\kms\ cloud and onto the CND (see Fig. \ref{fig:PV-diagrams}).

Broader and weaker emission is found in the CND, mostly within the NE Arm. The LSR velocities here are consistent with the NE Arm (50--80~\kms), and the structure is seemingly composed of multiple components. This is in contrast with the Southern Streamer, which appears consistent with a single strong peak as discussed in Section \ref{Dis:SiOCND}.

Weak emission is found in the northwestern portion of the CND just south of the linear filament. The emission is weaker here than in the NE Arm and the Southern Streamer but consistent with higher transitions of SiO observed previously. The velocity is the most redshifted, ranging between 60 and 90~\kms.

\subsection{Gridded SiO data}\label{subsec:resultsGridd}

\begin{figure}

\fig{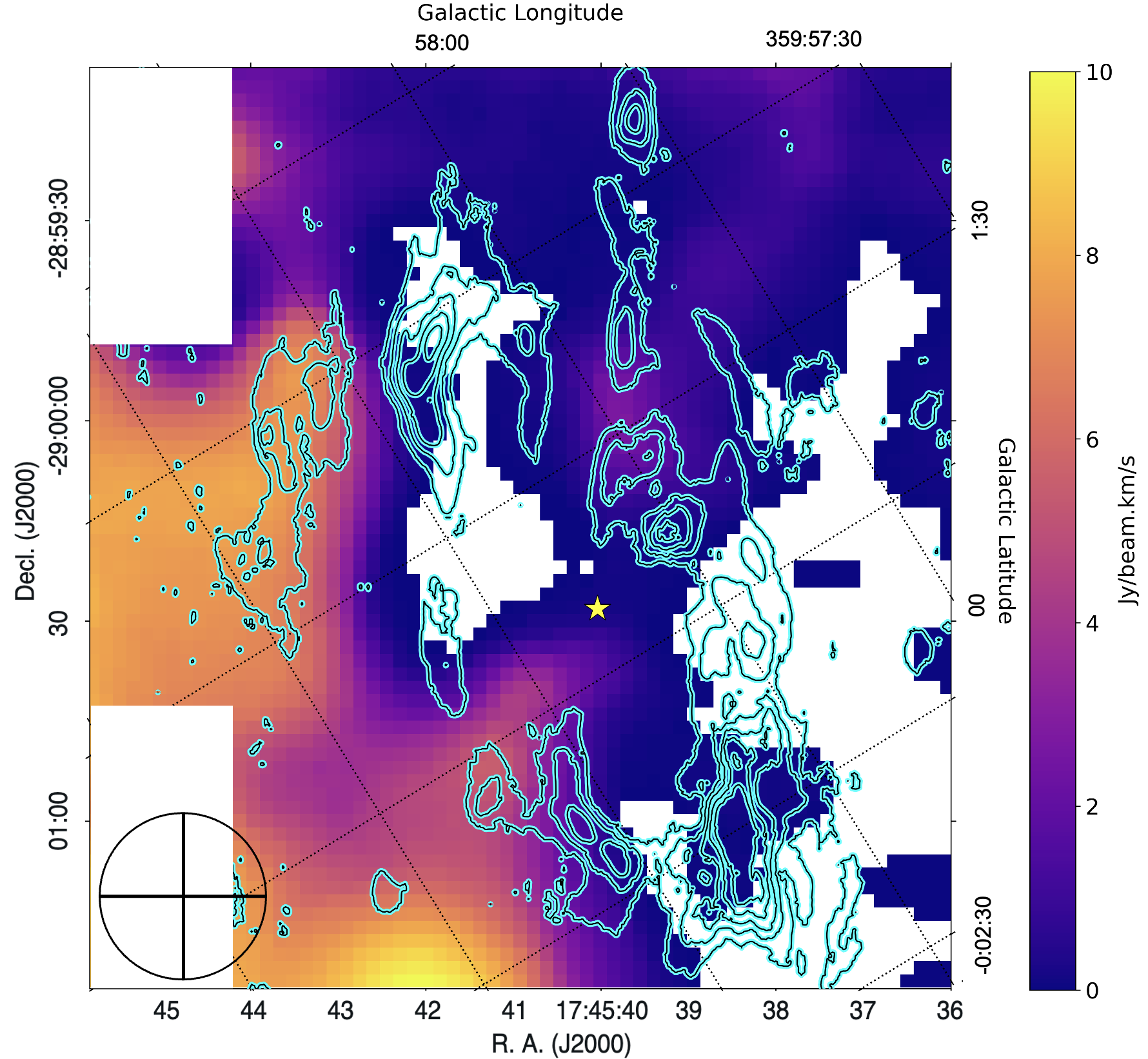}{1.0\textwidth}{}
\caption{Moment 0 map of the SiO with 25\arcsec\ resolution (shown bottom left). A mask of 2$\sigma$ was included to exclude false signal from the map, which is shown in white. Map displayed in equatorial coordinates with a galactic grid overlaid. Cyan contours are of HCN (1--0) to illustrate the structure of the CND. \sgra\ is represented with a star. The color bar goes from 0 to 10~\jyb~\kms.
The majority of the SiO emission is concentrated south-east of the CND and also in the NE Arm.
}
\label{fig:SiO_MOM0}
\end{figure}

We use the 110 pointings to create a spectral cube covering the whole region. Figure \ref{fig:SiO_MOM0} displays the SiO moment 0 map of the CND. Emission below 2~$\sigma$ is masked due to uncertain baselines discussed in Section~\ref{sec:Obs}. The sampling of the NE Arm extends farther than the CND map, so the top and bottom left edges (north of 28\degr 59\arcmin 30\arcsec, south of 29\degr 01\arcmin 00\arcsec, and east of 17:45:45) are masked. The strongest detection is 9.2~\jyb~\kms\ to the southeast of the CND near the edge of the field. The emission appearing on the edge of the gridded map results in uncertainty in the physical location of the peak emission, so the gridded data is used as a visual aid. A similar pattern of strong emission to the southeast and weak emission in the northwest appears in the spectral cube as in the postage stamp map, confirming that the cube interpolation worked properly. Of note is the emission east of the NE Arm that does not appear in the postage stamp map due to the lack of coverage, but now includes emission congruent with the edge of the Sgr A East region. A weak bridge of SiO emission over the southern extension of the CND to what may potentially be the East Arm of the minispiral appears in the integrated velocity map. A weak snaking structure also appears from the northwest portion of the CND towards the northwest. 

\begin{figure}
    \centering
    \includegraphics[width=\textwidth]{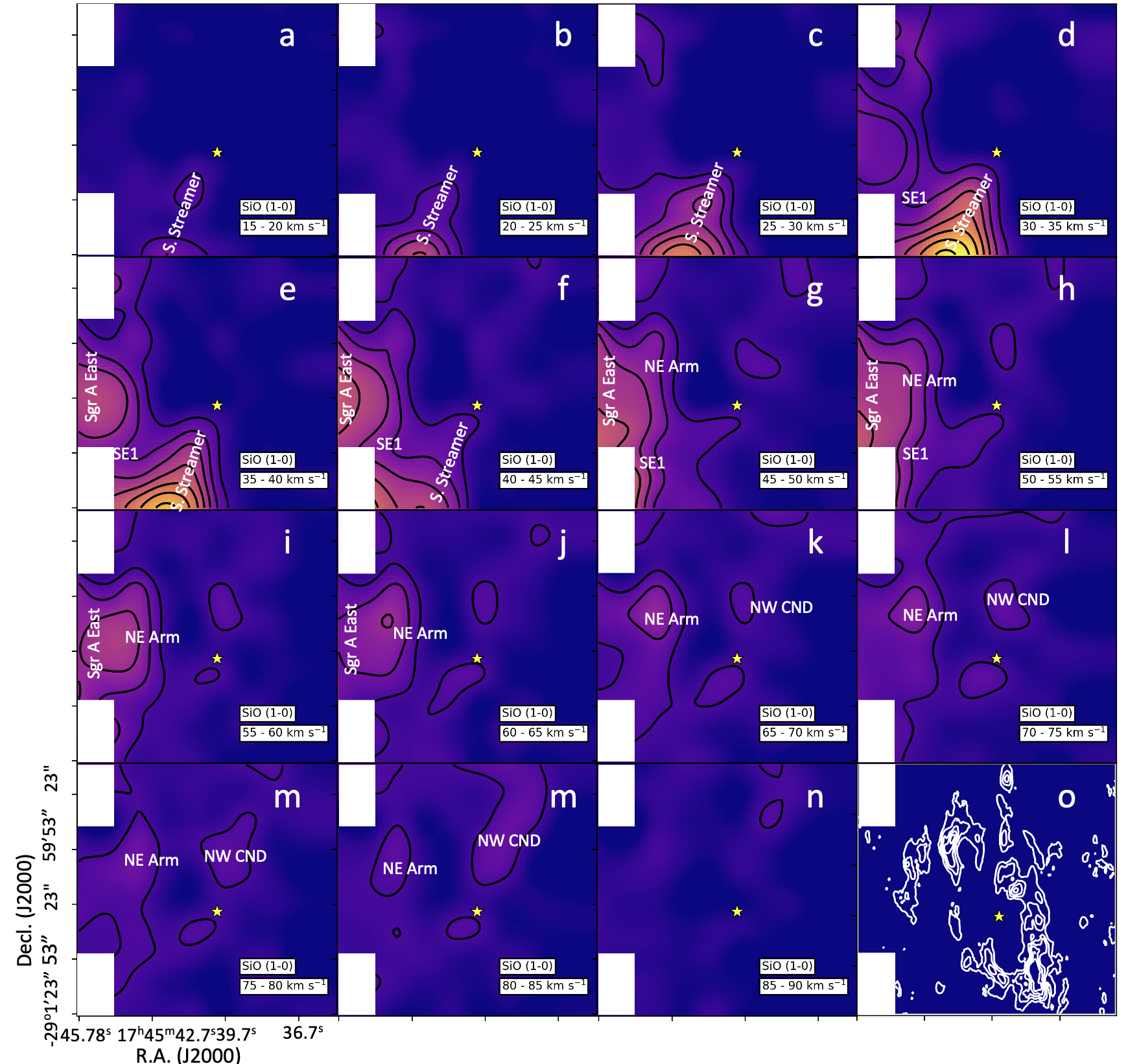}
    \caption{
    Interpolated map of the SiO (1--0) emission of the NE Arm averaged every 5~\kms. Panels labeled a-n start from 15-20~\kms\ and end at 85-90~\kms. Important features are labeled in white to help with ease of interpretation. The yellow star is the location of \sgra. The white contours in the bottom right panel (o) are to display the HCN (1--0) outline of the CND. Black contour levels start at 3~$\sigma$ and increase by 1~$\sigma$ (0.04~\jyb\kms).
    }
    \label{fig:sio_gradients}
\end{figure}

Figure \ref{fig:sio_gradients} shows the channel map average over 5~\kms\ starting at 15~\kms\ to 90~\kms. A small star indicating the location of \sgra\ appears in every panel for reference, and the HCN (1--0) contours of the CND are given in the bottom right panel for comparison. The strongest emission is once again found to the south east along the southern streamer from between 20 and 45~\kms. Emission appears east of the NE Arm between 30 and 60~\kms. Emission from the NE Arm appears between 40 and 75~\kms. The weak snake-like structure to the northwest appears weakly between 50 and 90~\kms and is strongest at 85~\kms.

\begin{figure}

\fig{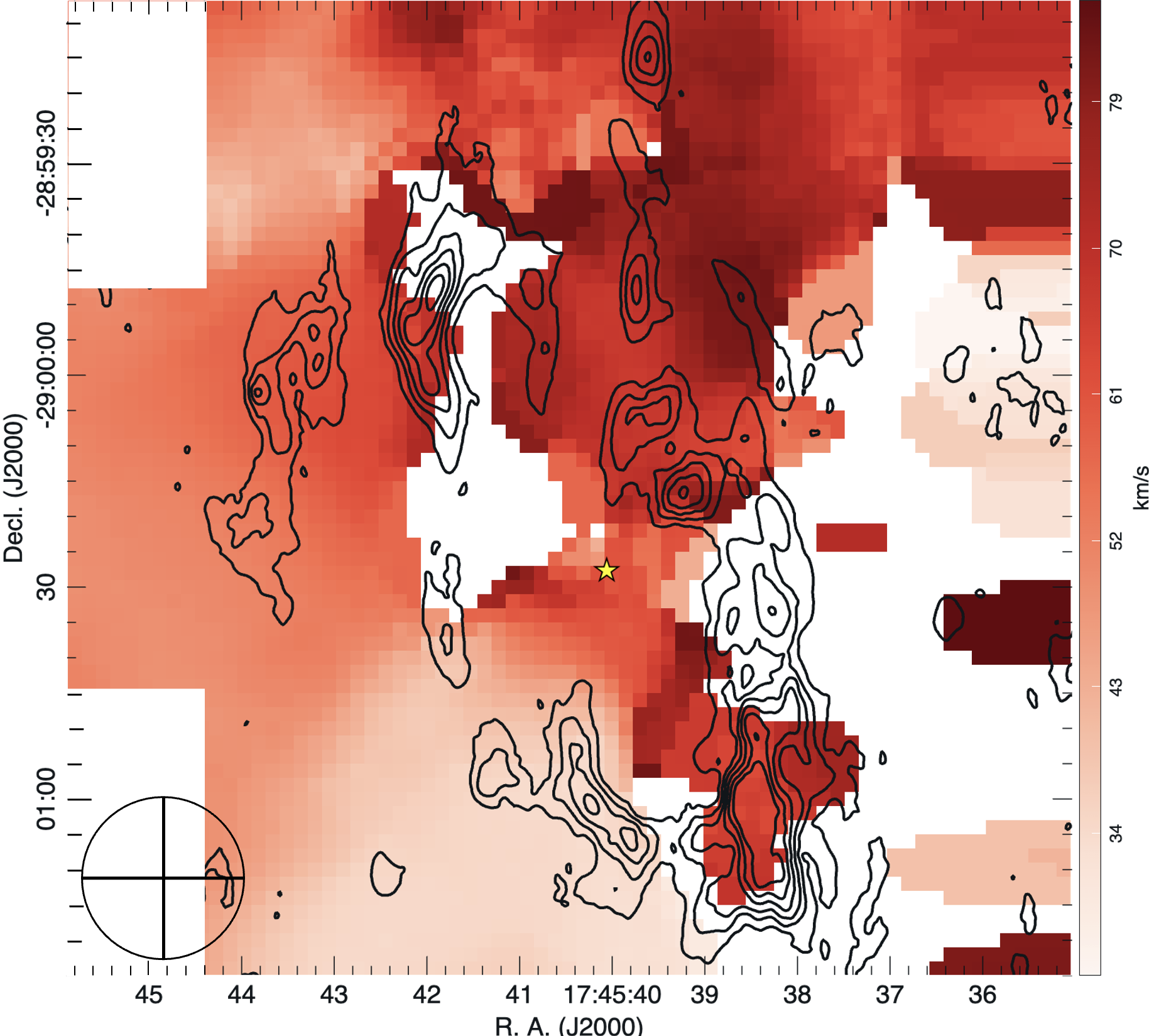}{1.0\textwidth}{}
\caption{Moment 1 map of the SiO (1--0) emission with 25\arcsec\ resolution (shown at the bottom left). Whited-out portions are where the emission fell below the 2~$\sigma$ cut-off. The map is displayed in equatorial coordinates. Black contours are of HCN (1--0) to illustrate the structure of the CND. The color bar is of velocity ranging from 28 to 83~\kms.
Note that all the SiO emission is redshifted, and the redshift is higher to the west than to the east.
}
\label{fig:SiO_MOM1}
\end{figure}

\begin{figure}

\fig{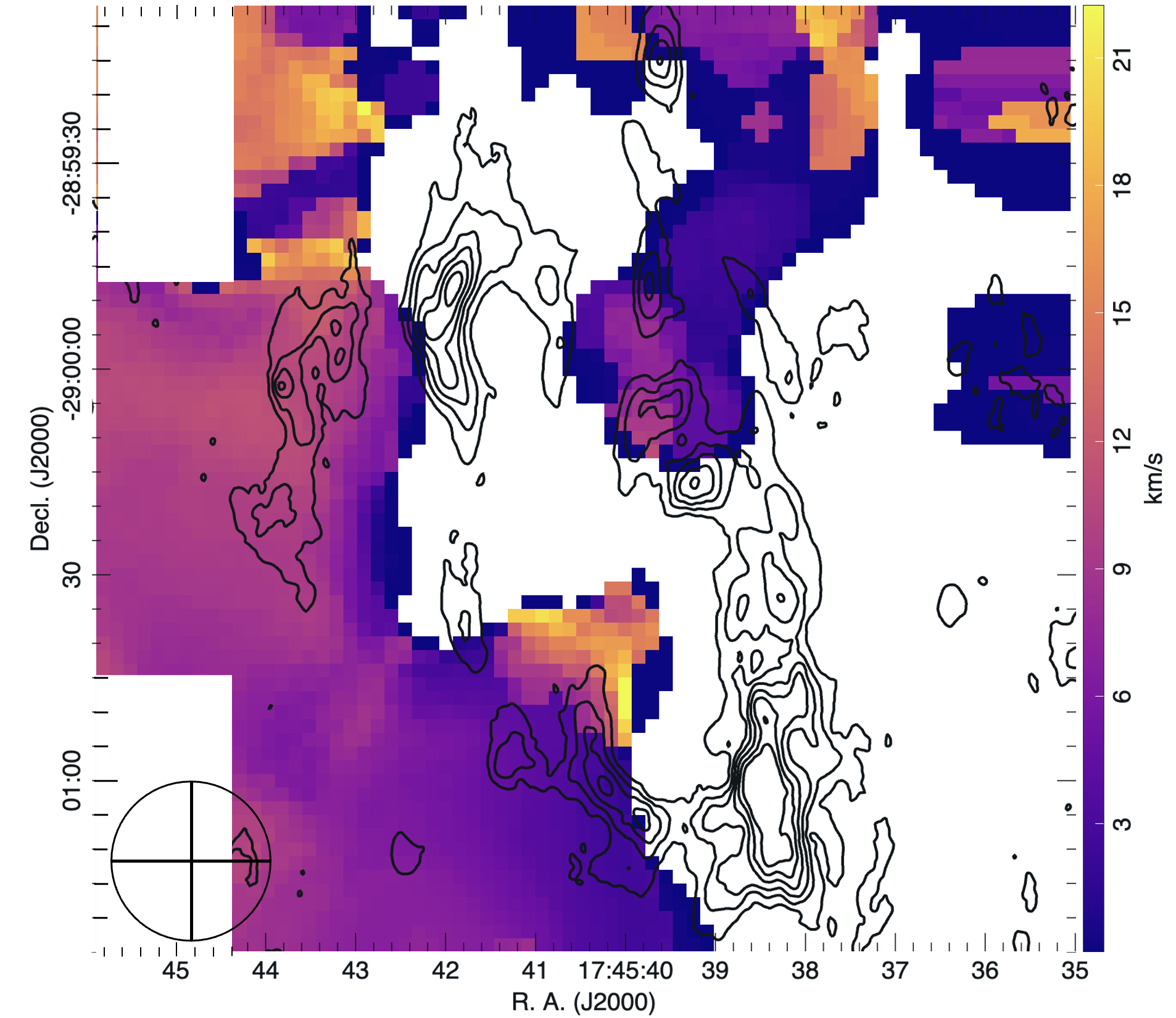}{1.0\textwidth}{}
\caption{Moment 2 map of the SiO (1--0) emission with 25\arcsec\ resolution (shown at the bottom left). Whited-out portions are where the emission fell below the 3~$\sigma$ cut-off. The map is displayed in equatorial coordinates. Black contours are of HCN (1--0) to illustrate the structure of the CND. The color bar is of velocity widths ranging from 0 to 22~\kms.}

\label{fig:SiO_MOM2}
\end{figure}

The SiO (1-0) emission measured in this work is generally redshifted. This contrasts with the CND as measured in other tracers, which typically exhibit a velocity range of $\pm150$~\kms\ \citep[e.g.][]{Montero-Castano2009GasDisk}. The moment 1 (intensity-weighted mean velocity) map in Figure \ref{fig:SiO_MOM1} demonstrates the level of the redshift. A cut-off value of 0.07~\jyb (\aproxi 2$\sigma$) is used to avoid including noise. The white areas are those masked by our threshold. The northwest region of the map moves faster than the southeast, agreeing with the postage stamp map. Figure \ref{fig:SiO_MOM2} shows the moment 2 (intensity-weighted dispersion of the coordinate) map of the SiO emission. The cut-off value of 0.11~\jyb\ (\aproxi 3$\sigma$) is used to avoid including noise. The white areas are those masked by our threshold. The velocity dispersion observed across the field lies between 0 and 22~\kms. 
The velocity of the SiO emission in the NW CND matches the velocities of the other molecular gas observed in the CND in that region. The discussion of the implications of the velocities seen in this map are saved for the discussion.

\begin{figure*}
    \vspace{0.1cm}
\centering
\begin{tabular}{cc}
    \includegraphics[width=0.40\textwidth]{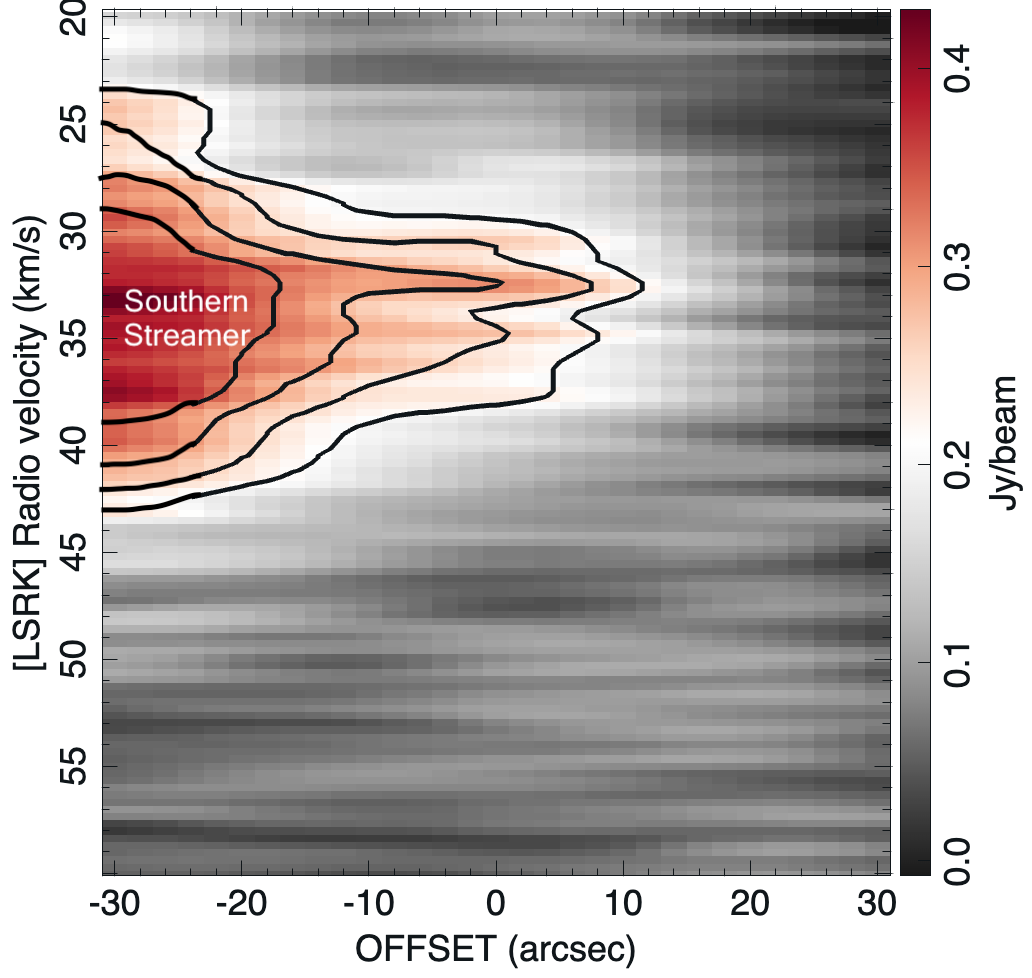}{a} &
    \includegraphics[width=0.40\textwidth]{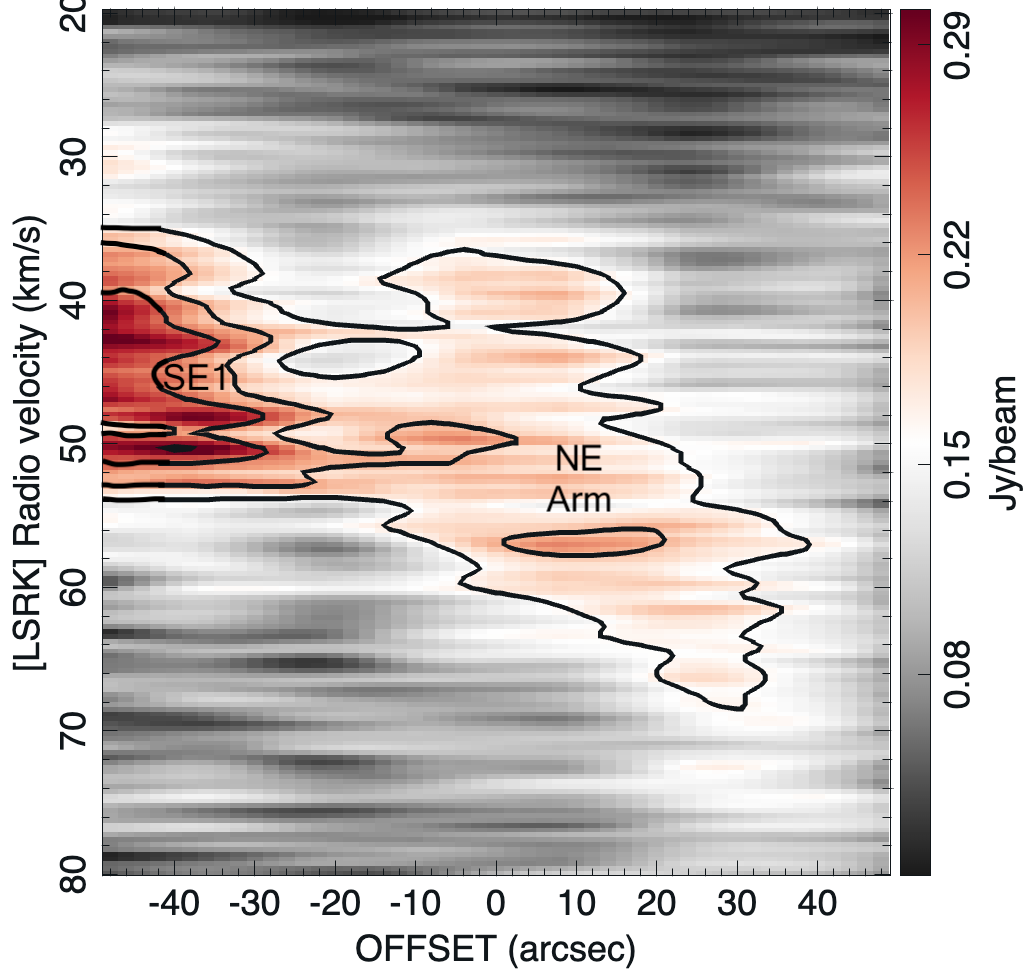}{b}
    \\
    \multicolumn{2}{c}{\includegraphics[width=0.83\textwidth]{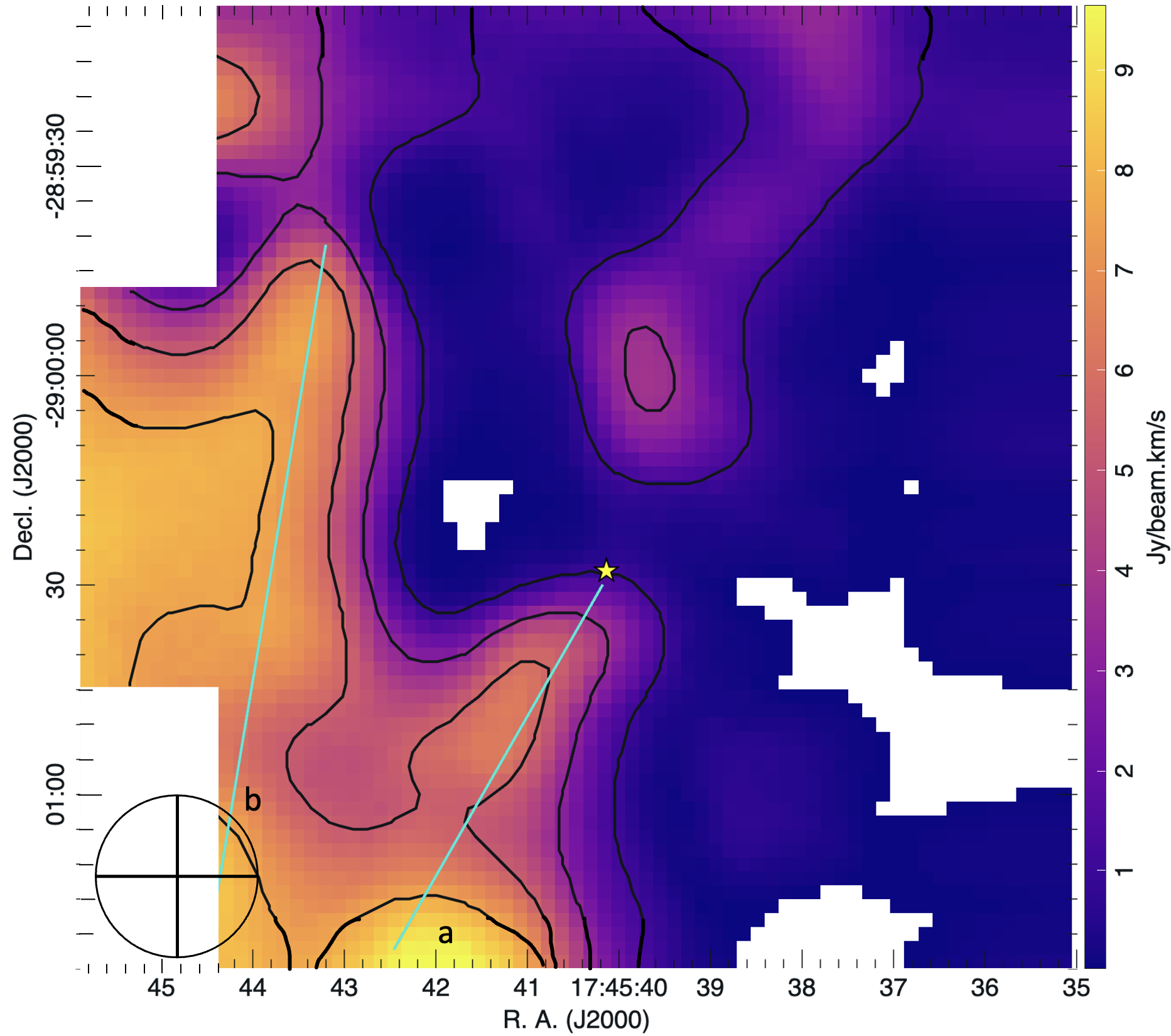}}
    \end{tabular}
    \caption{Top Left: PV diagram of the SiO emission along the Southern Streamer (a) starting from the southeast and moving northwest. Contours are fit starting at 3~$\sigma$ detection and increasing by 1~$\sigma$ per level for clarity. The line starts negative and moves positive. Top Right: PV diagram of the SiO emission along SE1 (b) starting from the southeast and moving northwest. Contours are fit starting at 3~$\sigma$ detection and increasing by 1~$\sigma$ per level for clarity. The line starts negative and moves positive. Bottom: Moment 0 intensity map of SiO in the CND with contours starting at 1~$\sigma$ per contour. The blue lines indicate the fitting used to make the PV diagrams above. The 25\arcsec\ resolution beam is shown on the bottom left.}
    \label{fig:PV-diagrams}
    
\end{figure*}

To explore the connection with the SiO emission towards the southeast with the potential relation to the southern streamer, SE1, and the CND, we create position-velocity (PV) diagrams. Figure \ref{fig:PV-diagrams} shows the position of two PV cuts along features believed to be the southern streamer (a) and SE1 (b) traced by ammonia reported by \citet{Mcgary2001INCLOUDS}. For each cut, the starting position is the most negative value, labeled by the letter in the figure. The 0 position is the center of the line. The contours start at 3~$\sigma$ and increase by 1~$\sigma$ per contour to help guide the eye along the emission. The contours in the map are of SiO and start at 3~$\sigma$ and increase by 1~$\sigma$ per contour. No emission outside the bounds of either graph is observed.

\section{Discussion} \label{sec:Discuss}
The following section details the SiO (1--0) large-scale emission dynamics of the region. The discussion is split into 3 subsections: Subsection \ref{Dis:SiOCND} focuses on emission in the CND. Subsection \ref{dis:sub:SIOStreamer} focuses on emission in the Southern Streamer. Subsection \ref{Dis:sub:Sgraeast} focuses on the SiO emission in Sgr A East.

\subsection{SiO in the CND} \label{Dis:SiOCND}
The majority of the SiO (1--0) emission found in the CND is central to the NE Arm and also weakly found in the NW near the linear filament. The velocities found in SiO (1--0) in these regions match those of the HCN (1--0) reported by \citet{Christopher2005HCNDISK}. Additionally, the SiO (1--0) peaks found in the NE Arm match those reported in the SiO (5--4) transition and the (2--1) transition observed using the Submillimeter Array (SMA) and the Combined Array for Research in Millimeter-wave Astronomy (CARMA) reported by \citet{Yusef-Zadeh2015SIGNATURESA}. In the Northwest CND, the SiO (1--0) emission also matches the velocities of the SiO (5--4) and (2--1) emission at the redder velocities \citep[see sources 25--27 in Table 5 of][for reference]{Yusef-Zadeh2015SIGNATURESA} but not at the bluer ($<$45~\kms) velocities (source 25). There are several explanations for not detecting the blueshifted emission from source 25: (i) the emission is compact, and the SiO(1--0) transition may be beam-diluted in our data -- the resolutions of the SMA and CARMA data are of $\sim 10$ to 30 times higher than the resolution of the interpolated GBT cube; (ii) the position of the emission falls at a location that is offset from the peak of GBT pointing \#131, and so its emission is suppressed because the map is under-sampled; or (iii) the shocked gas associated with source 25 is hotter than the other sources, such that the higher energy J-transitions are more populated than the lowest ones, making the SiO(1--0) more difficult to detect \citep{Gusdorf2008SiOWaves}. A combination of all these scenarios may be at work.
Only higher resolution (similar to CARMA and SMA) can determine to what extent beam dilution plays into the emission observed using the GBT.

Also of note is the lack of SiO (1--0) emission in the Southwest Lobe, the Northeast Lobe, the Southern Extension, and the Northeast Extension. These regions have SiO emission at higher transitions but lack the (1--0) transition. The lack of SiO (1--0) emission indicates that these regions likely host younger, hotter shocks due to the role the partition function plays in shocks \citep{Gusdorf2008SiOWaves}. In simulations, younger (more recent) shocks populate the higher transitions such as the (5--4), (4--3), and (3--2) with a peak in the (3--2) after about 3000 years as the shock cools. Simulations show that the 1–0 transition appears later as the J-levels repopulate. Additionally, the lower J-levels trace slower and cooler shocks, and high inclination angles along the line of sight ($>$ 60\degr) make the lower (J$<$4) transitions appear stronger \citep{Gusdorf2008SiOWaves}. A lack of emission from the lower transitions indicate that the shocks here are either (i) high velocity, (ii) hot and young, or (iii) a combination thereof when compared to the shocks observed in the NE Arm and the small portion of the NW CND. Observations of hotter shocks would support findings by \citet{James2021RevealingTransfer} and others that most of the CND is composed of hot, ionized gases that reach temperatures over 500~K.

\subsubsection{Star formation or cloud-cloud collision in the NE Arm?} \label{subsec:SForCC?}
The NE Arm differs greatly from the rest of the CND in SiO (1--0) emission. The NE Arm region of the CND sits slightly further from \sgra\ than the rest \citep[3~pc vs 0.5 to 2~pc respectively; i.e.][]{Christopher2005HCNDISK,Hsieh2021TheCenter}. Therefore, the environmental effects of the SMBH, such as radiation pressure and tidal shear, are reduced in the NE Arm compared to the CND as a whole. The broad (40--60~\kms) emission found here is strong compared to that found in the NW CND. Additionally, the spectra in the NE Arm display multiple narrow to broad (2--10~\kms) peaks indicative of multiple source components. The narrower components are seen in the NW CND, but the peaks there do not have components with FWHM $>$15~\kms. While some very broad ($<$20~\kms) emission exists in the NE Arm, the narrower (5-10~\kms) components indicate different dynamics happening in the gas here. Similar peaks with similar FWHM are found in the SiO (1--0) transition in the NE Arm as with the SiO (5--4) and SiO (2--1) transitions \citep{Yusef-Zadeh2015SIGNATURESA}. The peaks in the (2--1) transition appear stronger than the (1--0) by almost twice the intensity, but the (5--4) is slightly weaker than the observed (1--0). If beam dilution is once again playing a role in dampening the compact emission in the NE Arm, then the SiO (1--0) line would be even brighter than the (2--1) transition, implying the shocks traced by the SiO in the NE Arm are older than in the rest of the CND \citep{Gusdorf2008SiOWaves}. \citet{Yusef-Zadeh2015SIGNATURESA} claimed that the SiO emission, observed in the NE Arm with an interferometer at 3.6\arcsec\ resolution, are outflows driven by protostars. The low resolution (25\arcsec) GBT SiO (1–0) observations cannot resolve such a structure at the Galactic Center, and indeed would be severely diluted within the GBT beam, so we cannot confirm such a claim. However, the data display similar spectral emission in the overarching region, confirming that the (1–0) transition also appears in the region with similar velocity structure. Observing the (1–0) transition does not indicate the presence of protostellar outflows but does confirm the presence of shocks.

Additionally, when the SiO (1--0) observations are compared to the NH$_3$(3,3) observations from \citet{Mcgary2001INCLOUDS}, the observed strong emission from the NH$_3$(3,3) match the SiO centered on the NE Arm at velocities of 50~\kms\ in the south and 60~\kms\ in the north. The NH$_3$ is broad ($>10$~\kms) and winged here. Winged emission is often found in shocks associated with outflows \citep{Bally2016ProtostellarOutflows}, but there is no strong evidence for winged features in the SiO seen in the NE Arm in the GBT beam. The ambient NH$_3$ emission traces the faint streamer named SE1, and this streamer is believed to feed the CND. In the SiO (1--0), both peaks at the 50~\kms\ and 60~\kms\ exist in addition to more extended emission towards 70 to 80~\kms. Therefore, these shocks might be induced from the SE1 streamer colliding with the NE Arm, though the line widths of the emission are broader ($>4$~\kms) than currently expected with low-velocity shocks associated with cloud-cloud collisions \citep{Kim2023ARegions}. 

The broad, winged features point to outflow behavior found in star-forming regions \citep[e.g.,][]{Bally2016ProtostellarOutflows}, and there are other molecular signatures found in the NE Arm that trace dense molecular clouds that hint at the SiO (1--0) emission pointing to active high mass star formation.
These signposts include class I \meth\ masers that trace outflows and their associated kinematics \citep{Yusef-Zadeh2008MASSIVECENTER}, CS and \nh\ emission that traces dense molecular clouds that form protostellar cores \citep{Moser2017ApproachingA,Hsieh2021TheCenter,James2021RevealingTransfer}, and submm radio sources that trace protostars \citep{Merello2015TheObservations}. 

The NE Arm is a region where multiple clouds appear to produce a single structure in projection. Two clumps called the F and G clouds \citep{Christopher2005HCNDISK} interact to form the NE Arm with the F-cloud to the north and G-cloud to the south. A 1.87~$\mu$m infrared dark cloud (IRDC) \citep{Moser2017ApproachingA}, a known precursor to star formation \citep{Bergin2007ColdFormation}, traces a 10\arcsec\ region of sky on the southern portion of the G-cloud (see Figure \ref{fig:Galactic_Center_schematic} for reference). The IRDC at 1.87~$\mu$m traces the ``v-shaped" \nh\ (1--0) emission reported by \citet{Moser2017ApproachingA}, as well as the HCN (1--0) emission at the same velocity. The v-cloud is not a part of the CND and is believed to be a foreground cloud due to the velocities not matching that of the G-cloud (30~\kms\ vs 50~\kms, respectively). This has led many to conclude that there is no active star formation occurring in the NE Arm but a potential for some in the foreground IRDC. However, not all of the observed \nh\ emission reported by \citet{Moser2017ApproachingA} is a part of the v-cloud.

The reported \nh\ and the observed \meth\ emission have considerably broad FWHMs that average $\approx30$~\kms, more than five times wider than those of this work's SiO (1--0) measurements. The broad linewidths are likely due to the low spectral resolution of 20~\kms\ of the ALMA observations but also could indicate that the \nh\ emission is not tracing star formation, as is true for the Galactic disk. \nh\ is a known tracer of high density gas in star forming regions; \nh\ is found in the denser regions of cores than can be traced by CO due to its freeze resistance being greater than CO \citep{Jrgensen2020AstrochemistryStars}. 
\nh\ is also observed with multiple hyperfine transitions, making disentangling the individual transitions difficult. The wide spectral resolution of the \nh\ ALMA observations result in those transitions appearing as a single line in a single channel, resulting in further confusion. Though, like many molecules near the SMBH, another mechanism could be responsible for the creation of \nh, such as high cosmic ray ionization rates as seen in Sgr B2 \citep{Santa-Maria2021B2Hot}. Additionally, recent simulations produced by \citet{Barnes2024_sims_CCsigs} show that \nh\ is actually a better tracer of bridging features in cloud collisions in the Galactic Center, which is an important possibility to consider as some of the \nh\ emission reported by \citet{Moser2017ApproachingA} appears where the F-cloud and G-cloud potentially meet. The \nh\ emission near where the F-cloud and G-cloud meet has a central velocity of 49$\pm$7~\kms, which matches that of the G-cloud and higher J-transition SiO emission \citep{Yusef-Zadeh2015SIGNATURESA}. These two signatures together might indicate cloud-cloud collision as the mechanism for creating this emission. There are also observed 44- and 36~GHz class I \meth\ masers offset from said emission by about 0.1~pc, which could indicate signs of a protostellar outflow rather than cloud-cloud collision \citep{Pratap2008ClassFormation,McEwen2016CENTER,Cotton2016ADATA,Leurini2016PhysicalMasers}. Additionally, the G-cloud is home to an overdense CS clump \citep{Hsieh2021TheCenter} and a submm source \citep{Merello2015TheObservations}.
Finding \nh\ (1--0) emission in conjunction with a dense CS clump, a submm source, multiple transitions of class I \meth\ masers, and potential SiO outflows points to the early stages of star formation. If star formation is active in the CND, then the NE Arm would be the place to search.

The only way to determine whether star formation or cloud-cloud collisions are the driving force behind the SiO emission here is high resolution ($<1$\arcsec) observations of the SiO (1--0) line. If cloud-cloud collisions are the cause, then two velocity peaks will appear spatially co-located where the clouds are colliding. An additional bridge-like feature between the two peaks where the gas is interacting should also be present(see Figure 5 from \citet{Zeng2020Cloud-cloudClouds} for reference). If the emission is star formation, then the two peaks in velocity will be spatially separated in the form of an outflow with wing-like features. Additionally, we would expect to find a continuum source associated with driving the outflow.

\subsection{SiO in the Southern Streamer} \label{dis:sub:SIOStreamer}
The SiO (1--0) emission tied to the southern streamer is the strongest emission in the GBT data. There are high intensity ($>$~500~mJy), narrow ($<10$~\kms) peaks and also weaker features (see CND80 in Figure \ref{fig:SiO_CNDMAP}), indicative of shocks within the streamer \citep[i.e.][and references therein]{Kim2023ARegions}. The SiO (1--0) emission extends north along the streamer towards the CND but does not appear to interact with the CND. The weaker SiO emission in the southern portion of the CND does not match the blue-shifted velocities of the HCN. Therefore, the shocks in the southern streamer have no association with the CND. Additionally, the velocities do not match that of the gas in the east arm of the minispiral (V$_{LSR} \sim$ 100~\kms\ to 300~\kms).

When comparing the SiO line widths shown in Figure \ref{fig:PV-diagrams} with the observed ammonia dispersions reported by \citet[][see Figures 11 and 12]{Mcgary2001INCLOUDS}, the SiO traces the ammonia, both in location and in velocity dispersion space, which indicates that the ammonia and the SiO are tracing the same structures. The correspondence also appears in the moment 0 maps between the SiO and ammonia. The origin of the shocks in the southern streamer is difficult to determine. The velocities and velocity dispersions do not match those of the molecules tracing the CND nor the minispiral, which indicate that the SiO emission observed within the southern streamer is not due to interactions between the southern streamer and the CND. The shocks here are brighter than in the rest of the observed region, adding to the mystery of their origin. In the ambient dark clouds in the rest of the Galaxy, the fractional abundance of SiO is on the order of $10^{-12}$, and increases by 5 orders of magnitude in regions associated with shocks to $10^{-7}$ \citep{Martin-Pintado1992SiOOutflows,Jimenez-Serra2007ACluster}. The widespread SiO emission in the Galactic center has typical abundances of \aproxi$10^{-9}$ derived by \citet{Martin-Pintado1992SiOOutflows}. \citet{Amo-Baladron2011MappingA} discusses the possibility of non-shock related mechanisms for the increased abundance in the Galactic Center, especially in the vicinity of \sgra. They note the potential for X-ray and UV radiation producing an increased abundance in SiO, but point out that in observations of star forming regions within the Galactic Disk where X-ray and UV radiation are greater, there are no regions where the high energy radiation increase results in a significant increase in the SiO abundance. Instead, the observed abundance of SiO can be accounted for by the sputtering of grain mantles for c-shocks with velocities between 20 to 30~\kms. This agrees with simulations of c-shocks producing enhanced SiO abundances from \citet{Guillet2011ShocksShocks}. Therefore, the SiO is likely tracing large-scale shocks at the resolution of the observations in this paper.

The shocks produced by SE1 may be different than those observed in the southern streamer. SE1 travels north towards the NE Arm as discussed in Section \ref{subsec:SForCC?}. The position-velocity diagram along SE1 in Figure \ref{fig:PV-diagrams} shows that SE1 and the NE Arm are potentially interacting in velocity space, which is also seen from the ammonia emission presented by \citet{Mcgary2001INCLOUDS}. The ammonia and the SiO (1--0) emission trace the same features in PV-space and indicate that SE1 may be the source of the SiO emission observed in the NE Arm, implying that SE1 is interacting directly with the CND and feeding material onto the disk. However, the coarse resolution of the GBT cannot definitively confirm this, so further observations are necessary to confirm this interaction.

\subsection{SiO in Sgr A East} \label{Dis:sub:Sgraeast}

Strong, broad ($>$10~\kms) SiO (1--0) emission appears to the east of the CND from what is likely part of Sgr A East complex (see Figure \ref{fig:Galactic_Center_schematic} for reference). Nearby \meth\ masers tracing the edge of the expanding shell match the velocities measured by the SiO emission here \citep[35~\kms\ to 55~\kms;][]{McEwen2016CENTER}. 
The velocities of the SiO observed range from 30 to 60\kms, and peak around 45~\kms, which matches the masers observed in Sgr A East. The SiO shocks here are likely produced by the expanding supernova shell colliding with the ambient medium nearby as the velocities agree with the masers tracing the shockfront \citep[i.e.][]{McEwen2016CENTER}. 
The emission near Sgr A East appears to comprise multiple velocity components (see e.g. CND115 of Figure \ref{fig:SiO_CNDMAP} and the PV diagram along line b in Figure \ref{fig:PV-diagrams}).
This could be indicative of shocks tracing protostellar outflows in the NE Arm, or the expansion of Sgr A East as it interacts with the NE Arm.
There are a few narrow peaks (\aproxi4~\kms), which suggest that the shockfront is interacting with the ambient medium \citep{Draine1993TheoryShocks,Gusdorf2008SiOWaves}.
At the current resolution, it is not possible to completely disentangle emission coming from Sgr A East from that from the NE Arm; higher spatial resolution observations are needed to separate the emission.

\section{Summary and Conclusion}\label{sec:conclusion}
We presented an SiO (1--0) map of the central 5~pc of the Milky Way around the SMBH \sgra. Strong emission tracing the southern streamer leading from the 20~\kms\ cloud towards the CND, the NE Arm of the CND, and also Sgr A East is detected. A lack of SiO (1--0) emission is found in the central 2~pc around \sgra. The lack of SiO (1--0) emission in the CND suggests that the shocks in the inner portions of the CND are hotter than the outer portions, which would agree with results found in previous studies \citep[e.g.][]{James2021RevealingTransfer,Bryant2021TheCentre}. Additionally, the best location for active star formation, based on the properties of star-forming regions in the rest of the Galaxy, is in the NE Arm of the CND, offset by \aproxi3~pc.

Higher resolution observations ($<1$\arcsec) of the SiO (1--0) emission in the NE Arm are critical to determine if star formation is active in the CND, and the relative contributions of protostellar outflows and cloud-cloud collisions.

The summary of our main points are as follows:
\begin{enumerate}
    \item SiO (1--0) is abundant on large scales in the central 5~pc of the Galactic Center.
    \item SiO (1--0) traces kinematics in the Southern Streamer, NE Arm of the CND, NW CND, and also the edge of Sgr A East.
    \item Weak SiO emission traces the NW CND.
    \item The Southwest Lobe, the Northeast Lobe, the Southern Extension, and the Northeast Extension display a dearth of SiO (1--0) emission.
    \item Strong, broad peaks with multiple components are found in the entirety of the NE Arm, and these indicate a likelihood for active star formation in the region. Identifying regions of active star formation near \sgra\ allows for better models of how stars form near SMBH.
    \item Higher resolution observations of SiO (1--0) are necessary to determine the mechanism for the emission in the NE Arm of the CND, the Southern Streamer, Sgr A East, and NW CND.
    \item Strong emission is found in the Southern Streamer leading from the 20~\kms\ cloud to the CND. The emission here could indicate active cloud-cloud collisions within the streamer.

\end{enumerate}

\section{Acknowledgments}

We would like to thank the anonymous referee for their helpful insights and suggestions as their feedback made this paper significantly better. We also would like to thank Daniel Kabat and Caitlin Witt for their discussions, comments, and insights on the project. These data come from the NRAO archive project 07B-037. The National Radio Astronomy Observatory and Green Bank Observatory are facilities of the U.S. National Science Foundation operated under cooperative agreement by Associated Universities, Inc. JM is supported by an NSF Astronomy and Astrophysics Postdoctoral Fellowship under award AST-2401752.

\bibliographystyle{aasjournal} 
\bibliography{references,export-bibtex} 
\newpage

\end{document}